\documentstyle[prb,aps,multicol,eqsecnum,epsf]{revtex}
\begin{document}
\title{Field Theory on the von Neumann Lattice 
and the Quantized Hall Conductance of Bloch Electrons }
\draft
\author{K. Ishikawa, N. Maeda, T. Ochiai, and H. Suzuki}
\address{Department of Physics, Hokkaido University, Sapporo 060, Japan}
\date{\today}
\maketitle
\begin{abstract}
We construct useful sets of one-particle states 
 in the quantum Hall system based on 
the von Neumann lattice. 
Using the set of momentum  states, 
we develop a field-theoretical formalism
and apply the formalism to the  system subjected to a periodic potential.
The topological formula of the 
Hall conductance written by the winding number of  propagator 
is  generalized to  Bloch electrons.
The relation between the winding number  and the Chern number
is clarified.
\end{abstract}
\vskip20pt

\section{Introduction}

Current semiconductor technology
can produce a class of an electric modulation in a two-dimensional system.
In a two-dimensional electron system under a  magnetic field 
such a modulation causes quite interesting structure 
in various observables\cite{mod}.
It is also  used to examine various aspects of 
 quantum  Hall system\cite{aspects}.
The system subjected to a periodic  potential which is described by 
finite number of harmonics 
has been extensively studied by many authors\cite{hof,tknn,skg,cos}.
The system subjected to a cosine potential  is equivalent to 
the nearest-neighbor (NN) tight-binding model
with a magnetic flux. 
Hofstadter computed 
its  spectrum of butterfly-shape and discovered  a multi-fractal
structure\cite{hof}.
Other periodic potential may also have  a self-similar structure 
in a quantum region.
Experimental observation of such a  structure is a
challenging theme. 
The Hall conductance is a significant observable to extract physical 
properties 
of  such a system. 
Thouless {\it et al.} showed that in the system subjected to 
a periodic potential the Hall conductance is written  by 
the Chern number of the Bloch function in a gap region\cite{tknn,kohmoto}. 
In the cosine potential case the Chern number is 
given by a solution of the Diophantine equation. 
The spectra  and the Hall conductances in the next NN model
\cite{hk} 
and in  periodic potentials 
 which are  described by finite number of harmonics\cite{skg}  
are also obtained. 
However, the problem of  
a periodic short-range potential remains to be solved. 
Recently, a remarkable progress was made. 
That is, its spectrum in the lowest Landau level was computed with sufficient 
accuracy to see  self-similarity\cite{gb,imos}.
However, effects of higher Landau levels and Landau level mixing 
is not known well. In addition the Hall conductance of  subbands 
is unknown.  
A suitable base function is necessary for this calculation.
Many authors have used the Landau function or the eigenfunction of the 
angular momentum. 
The former one  extends in one direction of spatial coordinates  
 and localized in the other direction. 
This is suitable for 
the potential of  one-dimensional translational invariance.  
The latter one is  localized in the radial direction 
and  has a rotational invariance. 
This is suitable 
for a rotational invariant potential. 

The magnetic von Neumann lattice\cite{iimt,Neumann}
 is another basis in a magnetic field and  
can be constructed  independently of the gauge.
The basis consists of  direct products of 
coherent states  in the guiding center coordinate space 
and  harmonic oscillator eigenstates in the relative coordinate space. 
The von Neumann lattice basis is a set of localized states.  
It turns into a set of extended states by the Fourier transformation.
This basis has the following desirable properties:
(i)   A two-dimensional lattice translational invariance exists. 
(ii) Lattice structure of the von Neumann
           lattice can be changed according to a problem.  
(iii)A  modular invariance of the von Neumann lattice exists. 
Owing  to the first property, the Hilbert space of one-particle states 
is specified by the Landau level index and the lattice momentum.
 The second property is desirable 
in solving various periodic potential problems. 
The modular invariance is a  key to develop a field theoretical 
formalism based on the von Neumann lattice.
In our previous paper we showed advantages of the von Neumann lattice 
 by studying the spectrum of a periodic
short-range potential in the lowest Landau level\cite{imos}. 
A field theoretical formalism is necessary  when we study various 
interactions in a systematic way. 
In fact the field theoretical formalism based on the  von Neumann lattice 
allow us to  
prove the exactness of the quantized Hall conductance 
by a topological formula\cite{iimt,imatsu}. 
The topological formula is written by the winding number of the 
full propagator in the momentum space. 
The weak localization, QED effects, and electron-phonon interaction 
do not alter the quantized Hall conductance.
However, a periodic potential is not included in the proof.

In the present  paper we  study dynamical properties of the system 
subjected to a periodic potential using the von Neumann lattice. 
We first construct explicitly three sets of one-particle states 
based on the von Neumann lattice.
That is, the set of coherent states, the set of momentum states and 
the set of Wannier states. 
The coherent state is localized on a lattice site and is not orthogonal.
The momentum state is extended and orthogonal.
Its analytical expression in the spatial coordinate space was not known.
We give it in terms of the theta function for the first time.
The Wannier state is localized on a lattice site and is  orthogonal
\cite{ferr}.
Recently, this state was discussed by two groups\cite{rze,zak}.
However, their results are restricted to a rectangular von Neumann lattice. 
We study the analytical form of the Wannier state in a general 
von Neumann lattice.
Using the set of momentum states we develop a field theoretical formalism 
in a complete and strict manner. We pay attention to a peculiar property 
of the system in a magnetic field. In this system, we can not diagonalize 
both the current and the energy with respect to the Landau level index 
simultaneously. This imply that the usual relation 
$v_i={\partial \epsilon \over \partial p_i}$ does not hold in this system.
However, the gauge invariance allow us to prove  a similar relation 
as the Ward-Takahashi identity.
This identity is used to derive the topological formula of 
the Hall conductance. 
In a periodic potential system the spectrum consists of Landau subbands 
and the range of the lattice momentum 
becomes narrow according to the period of the potential.
The propagator has  additional index which reflects 
the subband structure. 
We prove that in this system the Hall conductance is written by 
the winding number of the  propagator. 
Furthermore, this winding number can be rewritten by the  Chern number 
of the eigenfunction.  
This  Chern number is the  same as the one 
obtained by Thouless {\it et al.}\cite{tknn}. 
As an example we calculate the Chern number of the system subjected to a 
periodic short-range potential. 
The result gives some insights to the Wigner crystal phase of the 
quantum Hall system.

The content of this paper is as follows.
We construct three sets of one-particle states based on 
the von Neumann lattice  in Sec.~II.
A field-theoretical formalism  based on the set of momentum states  
 is developed
in Sec.~III. The Ward-Takahashi identity and the topological  
formula of the Hall conductance written by the winding number of 
full propagator 
are shown.  
A periodic  potential and  defect are  studied 
using the formalism in Sec.~IV.
The Ward-Takahashi identity and the topological  formula
are generalized to a system subjected to a  periodic potentials in Sec~V.
The relation between the winding number and the Chern number  are clarified.
The Hall conductance of the system subjected to a periodic short-range 
potential are calculated with this formalism.
Summary and discussion are given in Sec.~VI.

\section{The von Neumann Lattice Basis}

In a two-dimensional system under a uniform perpendicular magnetic field $B$,
we introduce two sets of coordinates, {\it i.e.} the guiding
center coordinates  $(X,Y)$ and the relative coordinates $(\xi,\eta)$
\cite{kmh}:
\begin{eqnarray}
&\xi=(eB)^{-1}(p_y+eA_y),\ \ \eta=-(eB)^{-1}(p_x+eA_x),\\
&X=x-\xi,\ \ Y=y-\eta,
\end{eqnarray}
where $B=\partial_x A_y-\partial_y A_x$ and  $eB>0$.
Each set of coordinates satisfies the canonical commutation relation
\begin{equation}
[\xi,\eta]=-[X,Y]={a^2\over2\pi i},\ \ a=\sqrt{2\pi\hbar\over eB},
\end{equation}
and two sets of coordinates are commutative.
Using these operators, a
one-body Hamiltonian for a free charged particle is written in the form
\begin{equation}
\hat H_0={1\over 2}m\omega_c^2 (\xi^2+\eta^2),
\end{equation}
where  $\omega_c=eB/m$.
$\hat H_0$ is equivalent to the Hamiltonian of a harmonic oscillator
and the eigenvalue is solved as follows:
\begin{eqnarray}
\hat H_0\vert f_l\rangle=E_l\vert f_l\rangle,\
E_l=\hbar\omega_c(l+{1\over2}),\ \ l=0,1,2,\dots.
\end{eqnarray}
This energy level is called the Landau level.
Since $\hat H_0$ is independent of $X$ and $Y$, the phase space $(X,Y)$
corresponds to the degeneracy of the Landau level.
It is convenient to use the coherent state defined by
\begin{eqnarray}
&(X+iY)\vert\alpha_{mn}\rangle=z_{mn}\vert\alpha_{mn}\rangle,\\
&z_{mn}=(m\omega_x+n\omega_y)a,\nonumber
\end{eqnarray}
where $m,n$ are integers and $\omega_x,\omega_y$ are
complex numbers which satisfy
\begin{equation}
{\rm Im}[\omega_x^*\omega_y]=1.
\label{area}
\end{equation}
$z_{mn}$ is a point on the lattice site in the complex plane;
an area of the unit cell is $a^2$.
We call this lattice the magnetic von Neumann lattice.
Under the condition (\ref{area}), the completeness of the set
$\{\vert\alpha_{mn}\rangle\}$ is ensured\cite{pb}.
The coherent state $\vert\alpha_{mn}\rangle$ is constructed as follows:
\begin{eqnarray}
&\vert\alpha_{mn}\rangle=e^{i\pi(m+n+mn)+\sqrt{\pi}(A^\dagger 
{z_{mn}\over a}
-A{z^*_{mn}\over a})}\vert\alpha_{00}\rangle,\\
&A={\sqrt{\pi}\over a}(X+iY),\ \ [A,A^\dagger]=1.\nonumber
\end{eqnarray}
The coherent states are not orthogonal, that is,
\begin{equation}
\langle\alpha_{m+m',n+n'}\vert\alpha_{m',n'}\rangle =
e^{i\pi(m+n+mn)-{\pi\over2}{\vert z_{mn}\vert^2 \over a^2}}.
\label{orth}
\end{equation}
Thus, a translational invariance exists. 
The Hilbert space of one-particle states is spanned by the  state
$\vert f_l\otimes\alpha_{mn}\rangle$.
The expectation value of the position of this state is a coordinate
of the site of the magnetic von Neumann lattice and its mean square
deviation is $(l+1)a^2/2\pi$.
Hence, the state $\vert f_l\otimes\alpha_{mn}\rangle$ is a localized
state.
The wave function  of the state $|f_l\otimes\alpha_{mn}\rangle$ in the spatial 
coordinate space 
is given by 
\begin{equation}
\langle {\bf x}|f_l\otimes\alpha_{mn}\rangle =
{1\over a} \sqrt{{\pi^l\over l!}} ({z-z_{mn}\over a})^l
e^{-{\pi\over 2a^2}|z-z_{mn}|^2 
   -i\pi m{\tilde{y}\over a} +i\pi n{\tilde{x}\over a}
   +i\lambda({\bf x})}.
\label{alpwf}
\end{equation}
Here, $\tilde{x_i}=x_j W_{ji}^{-1}$ 
with the matrix $W$ defined by
\begin{eqnarray}
W=\left(
\begin{array}{cc}
{\rm Re}[\omega_x]&{\rm Im}[\omega_x]\\
{\rm Re}[\omega_y]&{\rm Im}[\omega_y]\\
\end{array}
\right),\ \ \det W=1.
\label{wm}
\end{eqnarray}
In terms of $\tilde{x},\tilde{y}$, $z=x+iy=\tilde{x}\omega_x
+\tilde{y}\omega_y$. Thus, a site of the von Neumann lattice
corresponds to integer $\tilde{x}/a,\tilde{y}/a$ value.
The function $\lambda({\bf x})$ represents a gauge degree of freedom 
of the eigenfunction.
For example $\lambda({\bf x})=0$  corresponds to  the symmetric gauge.
\begin{figure}[h]
\centerline{
\epsfysize=2in\epsffile{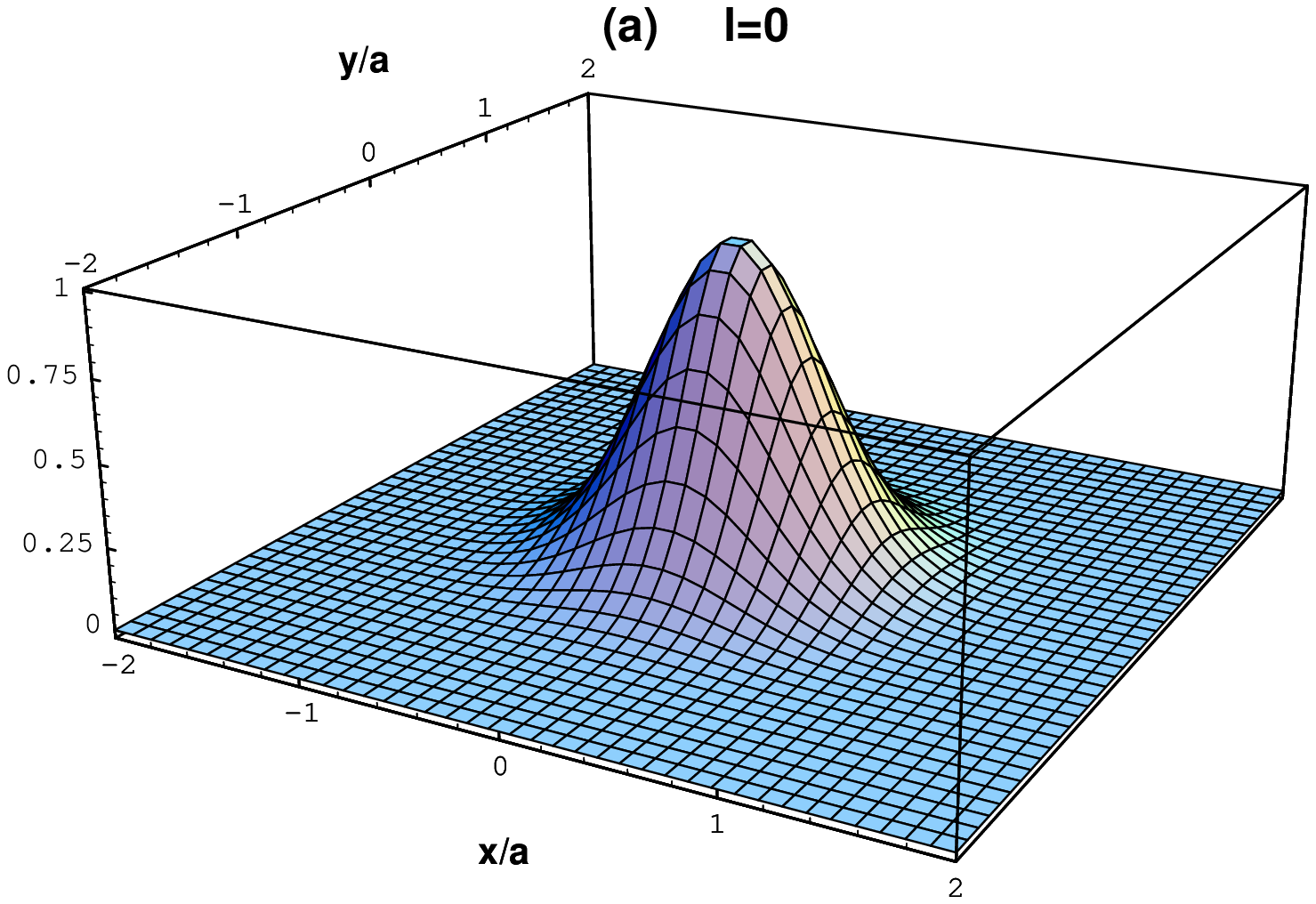}
\epsfysize=2in\epsffile{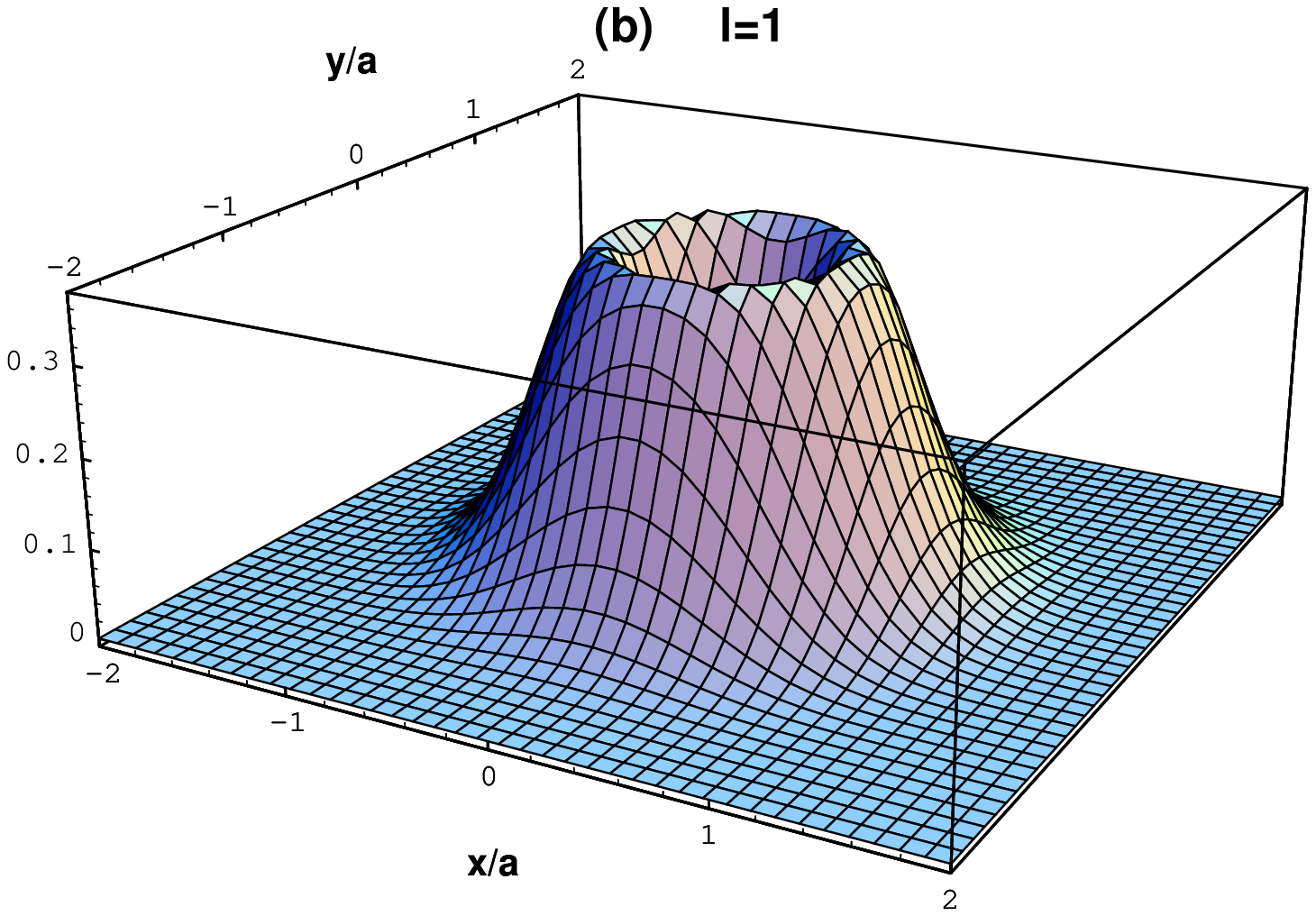}}
\caption{The probability density
$a^2 |\langle x|f_l \otimes \alpha_{00}\rangle|^2$ with $l=0,1$
for the square von Neumann lattice.}
\end{figure}
\noindent
In Fig.1 we show the probability density 
$a^2 |\langle x|f_l \otimes \alpha_{00}\rangle|^2$  with $l=0,1$ 
for the square von Neumann lattice. 
Localization around ${\bf x}={\bf 0}$ is clearly seen in this figure.
Since the probability density is a function of $\tilde{x}-ma$ and 
$\tilde{y}-na$, the correspondent figure  of $|f_l\otimes\alpha_{mn}\rangle$
is obtained by sliding the coordinate as 
$(\tilde{x},\tilde{y})\to (\tilde{x}-am,\tilde{y}-an)$.

As Eq.~(\ref{orth}) has a translational invariance,
an orthogonal basis can be obtained in the momentum space.
Fourier transformed states denoted by
\begin{equation}
\vert\alpha_{{\bf p}}\rangle=\sum_{m,n}e^{ip_xm+ip_yn}
\vert\alpha_{m,n}\rangle,
\end{equation}
are orthogonal, that is,
\begin{equation}
\langle\alpha_{{\bf p}}\vert\alpha_{{\bf p}'}\rangle=
\alpha({\bf p})\sum_{N}(2\pi)^2\delta({\bf p}-{\bf p}'
-2\pi{\bf N}).
\end{equation}
Here, ${\bf N}=(N_x,N_y)$ is a vector with  integer values 
and ${\bf p}=(p_x,p_y)$ is a momentum in the Brillouin zone (BZ), that is, 
$|p_x|,|p_y|\le\pi$. 
The function $\alpha({\bf p})$ is the Fourier transform of
Eq.~(\ref{orth}) and calculated by using the Poisson resummation
formula as follows:
\begin{eqnarray}
\alpha({\bf p})&=&\beta({\bf p})^*\beta({\bf p}),
\label{alpha}\\
\beta({\bf p})&=&\left(2{\rm Im}\tau\right)^{1\over4}
e^{i{\tau\over4\pi}p_y^2}
\vartheta_1({p_x+\tau p_y\over2\pi}\vert\tau),
\label{beta}
\end{eqnarray}
where $\vartheta_1(z\vert\tau)$ is a theta function and
the moduli of the von Neumann lattice is defined by 
$\tau=-\omega_x/ \omega_y$.
The magnetic von Neumann lattice 
is parameterized by $\tau$.
To indicate the dependence on $\tau$, we sometimes use a notation 
such as 
$\beta({\bf p}\vert\tau)$. 
For $\tau=i$, the von Neumann lattice becomes a square lattice.
For $\tau=e^{i2\pi/3}$,
it becomes a triangular lattice.
Some properties of the above functions are presented in Appendix A.
Whereas $\alpha({\bf p})$ satisfies the periodic boundary condition, 
$\beta({\bf p})$ obeys a nontrivial boundary condition
\begin{equation}
\beta({\bf p}+2\pi{\bf N})=e^{i\phi(p,N)}\beta({\bf p}),
\end{equation}
where
$\phi(p,N)=\pi(N_x+N_y)-N_y p_x$.
We can define the  orthogonal state  which is normalized with 
$\delta$-function as follows:
\begin{equation}
|\beta_{{\bf p}}\rangle={\vert\alpha_{{\bf p}}
\rangle\over\beta({\bf p})e^{i\chi({\bf p})}}.
\label{albe}
\end{equation}
The function $\chi({\bf p})$ reflects an ambiguity with respect to 
the normalization.  This  ambiguity implies another gauge symmetry 
of this system. Later, we identify this symmetry as the  gauge symmetry 
of constant magnetic field in BZ. 
It should be noted that the state $\vert\alpha_{{\bf 0}}\rangle$ is
a null state,  that is, $\sum_{m,n}\vert\alpha_{mn}\rangle=0$\cite{pb},
because $\beta(0)=0$.
Since $\vert\alpha_{{\bf 0}}\rangle/\beta(0)$ is indeterminate
in Eq.~(\ref{albe}), $|\beta_{{\bf 0}}\rangle$ is defined by
$\lim_{{\bf p}\rightarrow0}|\beta_{{\bf p}}\rangle$.
The Hilbert space of one-particle states is also spanned by the state 
$\vert f_l\otimes\beta_{{\bf p}}\rangle$.
We call the state $\vert f_l\otimes\beta_{{\bf p}}\rangle$ 
the momentum state of the von Neumann lattice.
The wave function of the state $\vert f_l\otimes\beta_{{\bf p}}\rangle$ 
in the spatial coordinate space is given by 
\begin{equation}
\langle {\bf x}| f_l\otimes\beta_{{\bf p}}\rangle =
{e^{i\lambda({\bf x})-i\chi({\bf p})}\over a}
\sqrt{\pi^l \over l!}({a\over 2\pi})^l 
(-2\partial_{z^*}+{\pi \over a^2}z)^l
\left(\beta^* (p_x-2\pi{\tilde{y}\over a},p_y+2\pi{\tilde{x}\over a})
e^{ip_y{{\tilde y}\over a} +i\pi{\tilde{x}\tilde{y}\over a^2}} \right).
\label{momentum}
\end{equation}
The probability density 
 $|\langle {\bf x}|f_l\otimes \beta_{{\bf p}}\rangle |^2$  
is invariant under the translation
$(\tilde x,\tilde y)\to (\tilde x +aN_x,\tilde y +aN_y)$ with integer 
$N_x,N_y$.
Thus, the momentum state is an extended state.
\begin{figure}[h]
\centerline{
\epsfysize=2in\epsffile{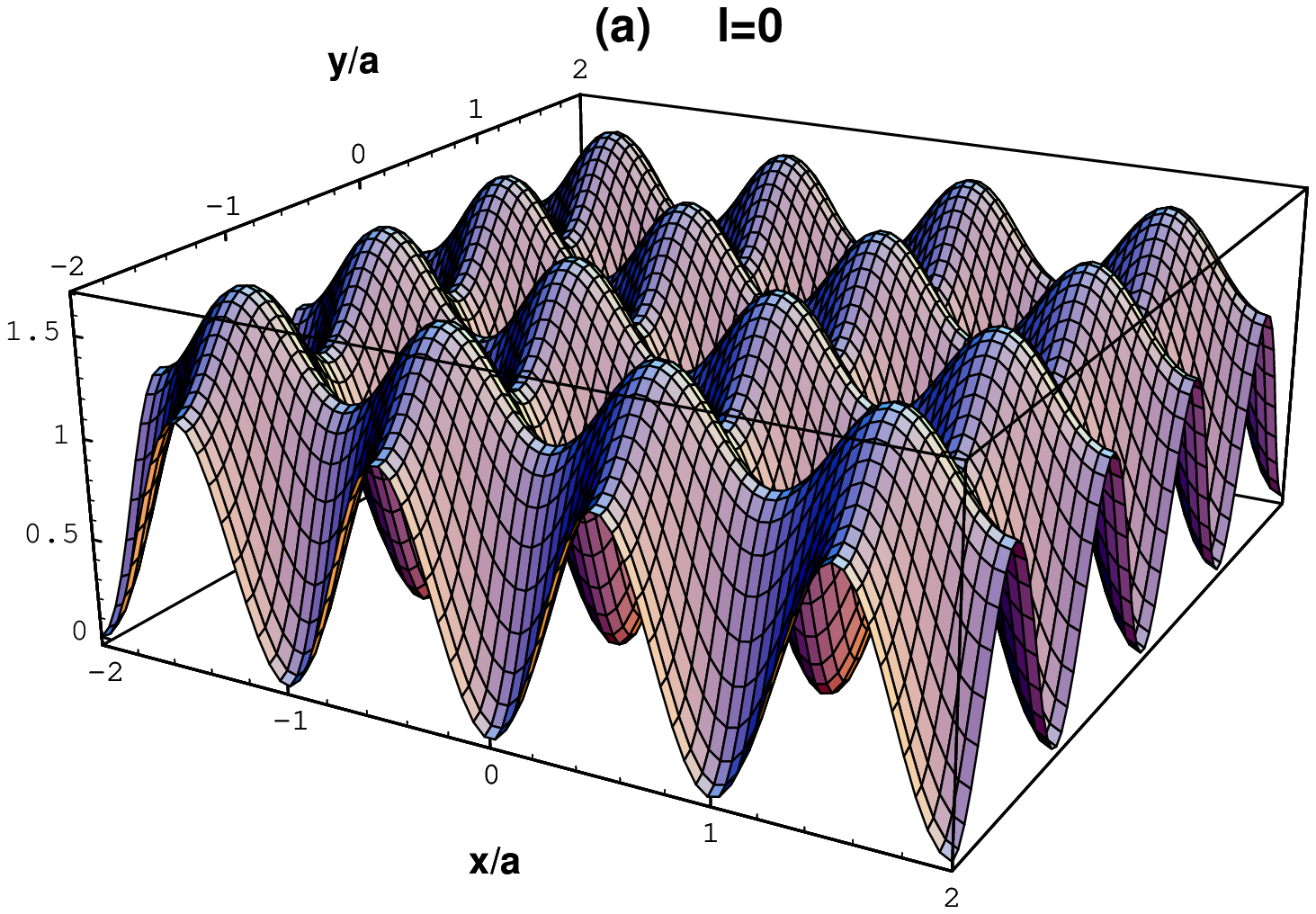}
\epsfysize=2in\epsffile{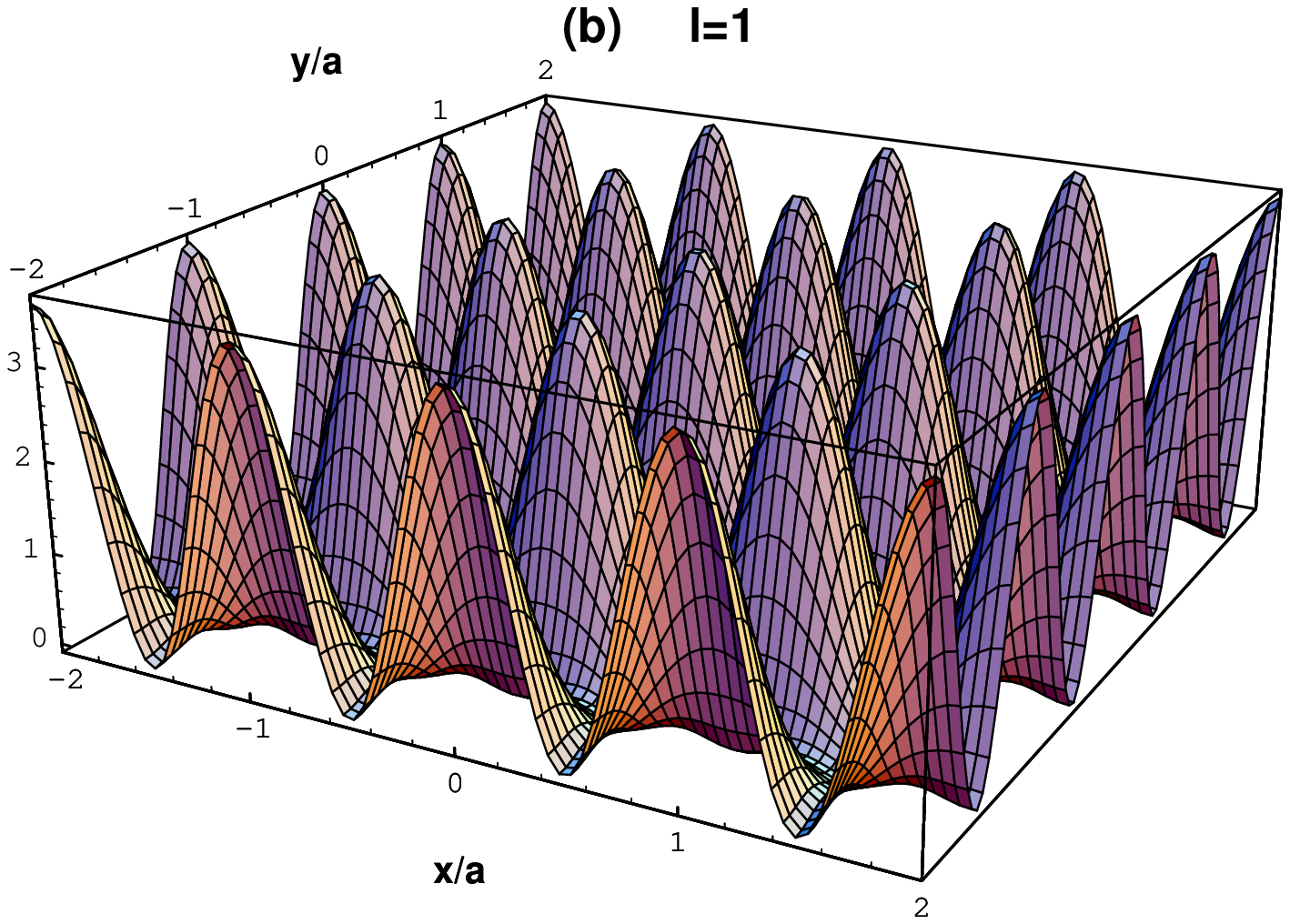}}
\caption{The probability density
$a^2 |\langle x|f_l \otimes \beta_{{\bf 0}}\rangle|^2$ with $l=0,1$
for the square von Neumann lattice.}
\end{figure}
\noindent
In Fig.2, we show the probability density for the lowest two Landau 
levels.
In Fig.2  the von Neumann lattice is the square lattice  and we choose 
${\bf p}={\bf 0}$. 
The probability density with a non-zero momentum ${\bf p}$ is
obtained by sliding the coordinate  as
$(\tilde x,\tilde y)\rightarrow(\tilde x+ap_y/2\pi,\tilde y-ap_x/
2\pi)$.

Since  $|\beta_{\bf p}\rangle$ is  orthogonal and 
normalized with $\delta$-function, 
its inverse Fourier transformation leads to an orthonormal state.
That is, the state defined by 
\begin{equation}
|\beta_{mn}\rangle =\int_{{\rm BZ}}{d^2 p\over (2\pi)^2}
e^{-ip_xm-ip_yn}|\beta_{{\bf p}}\rangle
\end{equation}
satisfies 
\begin{equation}
\langle \beta_{mn}|\beta_{m'n'}\rangle =\delta_{mm'}\delta_{nn'}.
\end{equation}
The state $|\beta_{mn}\rangle$ 
 is completely different from  the coherent state
$\vert\alpha_{mn}\rangle$ due to the normalization factor 
$\beta ({\bf p})e^{i\chi({\bf p})}$. 
The relation between the coherent state and the orthonormal state is given by 
\begin{equation}
|\beta_{mn}\rangle =\sum_{m'n'}G(m-m',n-n')|\alpha_{m'n'}\rangle ,
\end{equation}
where $G(m,n)$ is the inverse Fourier transformation of 
$1/\beta({\bf p})e^{i\chi({\bf p})}$.
It has a long tail proportional to $(m^2+n^2)^{-1/2}$.
The Hilbert space of one-particle states is also spanned by the state 
$|f_l\otimes\beta_{mn}\rangle$ which is  an orthonormal localized state
and has the center around  $z=z_{mn}$ in the spatial coordinate space. 
However, it also has a long tail away from the center. 
We call the state  $|f_l\otimes\beta_{mn}\rangle$ 
 the Wannier state of the von Neumann lattice.
The wave function of the state $|f_l\otimes\beta_{mn}\rangle$ in the spatial 
coordinate space depends  strongly  on the phase $\chi({\bf p})$.
A simple  form is obtained if we choose   
$\chi({\bf p})=p_xp_y/2\pi +(p_y+\pi)/2$.
In this case the wave function becomes  
\begin{eqnarray}
\langle {\bf x}|f_l\otimes\beta_{mn}\rangle &=&
-{e^{i\lambda({\bf x})}\over a}
\sqrt{\pi^l \over l!}({a\over 2\pi})^l 
(-2\partial_{z^*}+{\pi \over a^2}z)^l \nonumber \\
& & \times (2{\rm Im}\tau)^{{1\over 4}}\sqrt{{1\over i\tau^*}}
e^{-i\pi\tau^*({\tilde{x}\over a}-m)^2
-2\pi im {\tilde{y}\over a}+i\pi (m+n+{1\over 2})+
i\pi{\tilde{x}\tilde{y}\over a^2}} 
 \int_{-\pi}^{\pi}{dp_x \over 2\pi}e^{{i\pi\over \tau^*}
({p_x \over 2\pi}-{\tilde{y}\over a}+n+{1\over 2}+\tau^* 
({\tilde{x}\over a}-m))^2}. 
\label{wannier}
\end{eqnarray}
The similar result  was obtained by Zak for 
a rectangular von Neumann lattice in the lowest Landau level\cite{zak}.   
In Fig.3 we show the probability density 
$a^2 \langle {\bf x}|f_l\otimes\beta_{00}\rangle$ with $l=0,1$ for 
the square von Neumann lattice.  

\begin{figure}[h]
\centerline{
\epsfysize=2in\epsffile{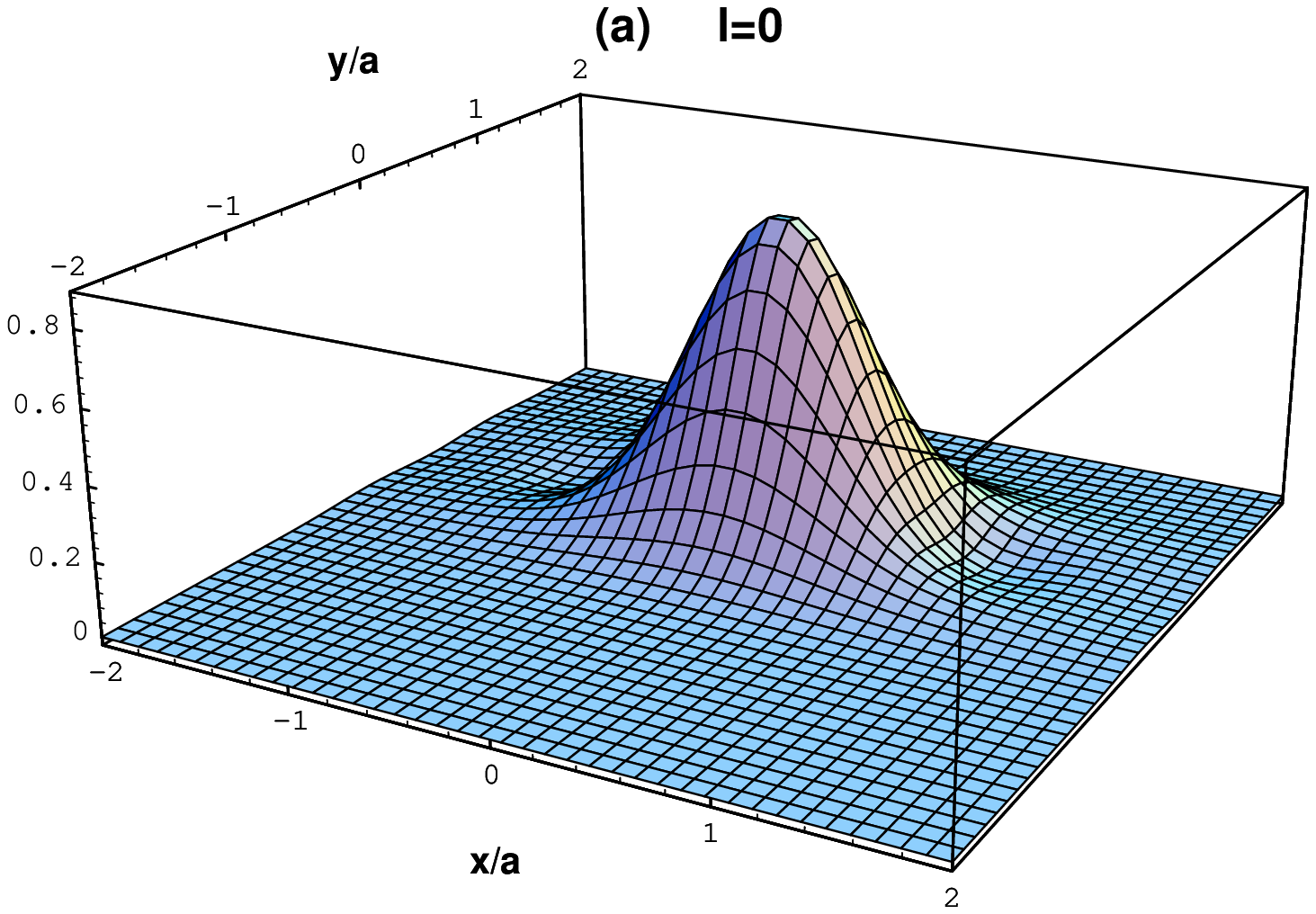}
\epsfysize=2in\epsffile{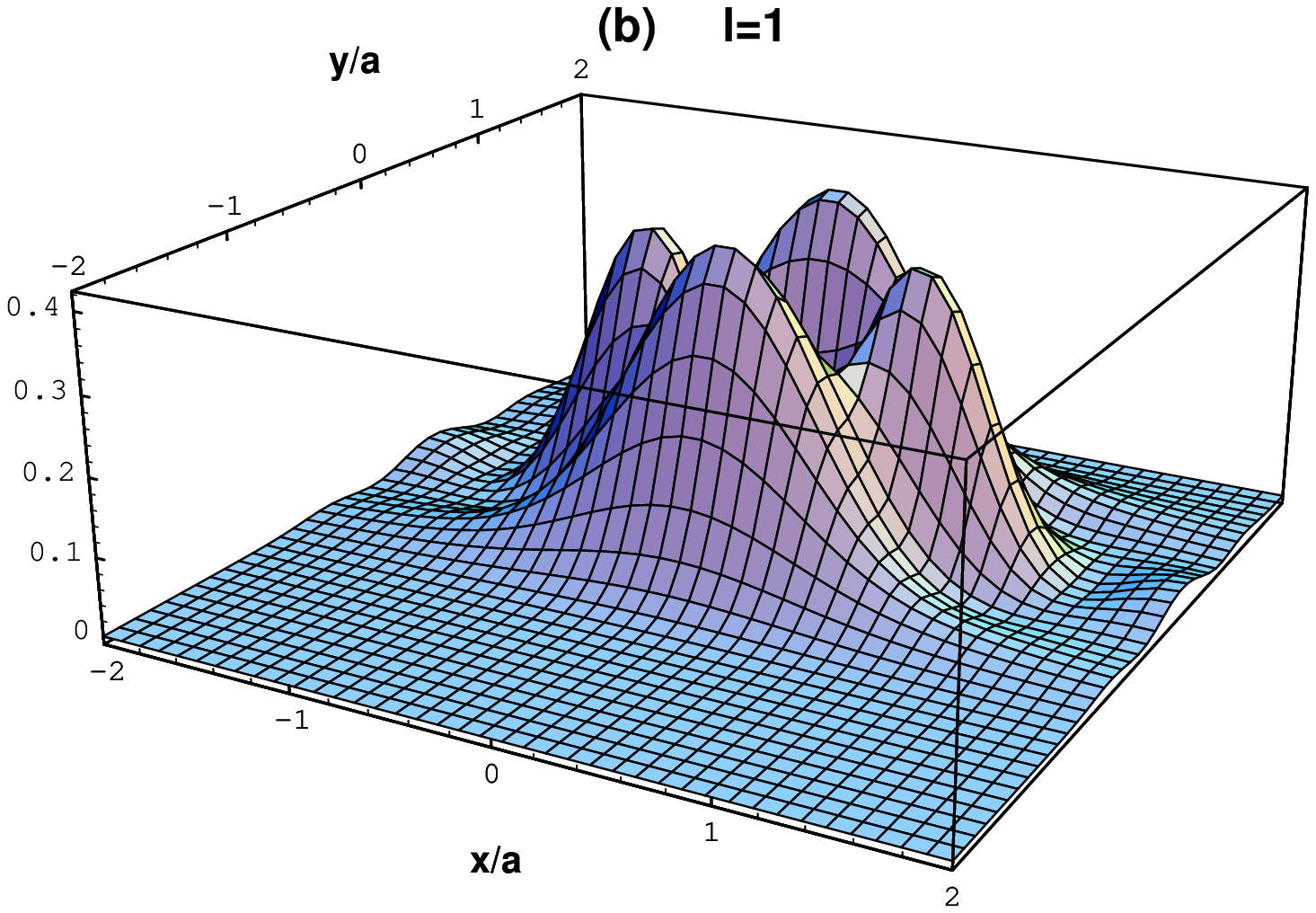}}
\caption{The probability density
$a^2 |\langle x|f_l \otimes \beta_{00}\rangle|^2$ with $l=0,1$
for the square von Neumann lattice.}
\end{figure}
\noindent
In Fig.3 we show the probability density 
$a^2 \langle {\bf x}|f_l\otimes\beta_{00}\rangle$ with $l=0,1$ for 
the square von Neumann lattice.  
Asymmetry between  $x$ and $y$-direction is owing to the phase choice.
The center of the wave function 
$\langle {\bf x}|f_l\otimes\beta_{00}\rangle$ 
is at $(\tilde{x},\tilde{y})=(0,1/2)a$ in contrast to a naive expectation
that it is at  $(\tilde{x},\tilde{y})=(0,0)$. This is not restricted to 
the case $\beta_{00}$. Since the probability density 
$|\langle {\bf x}|f_l\otimes\beta_{mn}\rangle|^2$ is 
a function of $\tilde{x}-am,\tilde{y}-an$, the slide of the center 
occurs for all $\beta_{mn}$.
We should note that the Wannier state found  by Rashba {\it et al.} 
 is slightly different from ours\cite{rze}.
They used  $|\beta ({\bf p})|$  
as the normalization factor.  
As a consequence, the probability density of the wave function   
$\langle {\bf x}|f_0\otimes \beta_{00}\rangle$ for the square von 
Neumann lattice has a center at $z=0$ and is rather symmetric than ours. 
Their  wave function  behaves as $1/r^2$  for large $r$ which 
is the  critical behavior in the  Thouless's criterion\cite{thoul}. 
Our wave function  behaves as $1/x$ for large $x$ and is exponentially dumped 
for large $y$. This behavior is also the Thouless's critical behavior.

\section{Field Theoretical Formalism and Topological Formula 
of Hall Conductance}

In the preceding section 
we obtain  three sets of one-particle states based on 
 the von Neumann lattice, that is, 
the coherent  state 
$|f_l\otimes\alpha_{mn}\rangle$, the momentum state
$|f_l\otimes\beta_{{\bf p}}\rangle$ and the Wannier state 
$|f_l\otimes\beta_{mn}\rangle$. A field theoretical formalism based on the 
coherent state  was developed in \cite{iimt}.
A formalism based on the momentum state  
or on the Wannier state  was  partially used in \cite{im}. 
Here we develop the field theoretical formalism based on the momentum state  
in a complete and strict  manner for later convenience. 
From now on, we denote $\vert f_l\otimes\beta_{{\bf p}}\rangle$ 
as $|l,{\bf p}\rangle$ and choose $\chi({\bf p})=0$. 
The reason  to prefer our choice ($\chi({\bf p})=0$) is 
 simpleness of the form for the density operator 
$\rho ({\bf k})$ discussed later. 
If we use the definition of \cite{rze}, $\rho ({\bf k})$ becomes a
complicated form.

We expand the electron field operator in the form
\begin{equation}
\psi({\bf x})=\int_{{\rm BZ}}{d^2p\over(2\pi)^2}\sum_{l=0}
^{\infty}b_l({\bf p})\langle {\bf x}|l,{\bf p}\rangle.
\label{field}
\end{equation}
$b_l({\bf p})$ satisfies the anti-commuting relation
\begin{equation}
\{b_{l}({\bf p}),b^\dagger_{l'}({\bf p}')\}=\delta_{l,l'}
\sum_N (2\pi)^2\delta({\bf p}-{\bf p}'-2\pi{\bf N})e^{
i\phi(p',N)},
\end{equation}
and the same boundary condition as $\beta ({\bf p})$.
$b^\dagger_l$ and $b_l$ are creation and annihilation operators
which operate on the many-body states.
The free Hamiltonian is given by 
\begin{equation}
{\cal H}_0=\int d^2 x \psi^\dagger ({\bf x})\hat{H}_0 \psi({\bf x})
=\sum_{l}\int_{{\rm BZ}}{d^2 p\over (2\pi)^2}E_l
 b^\dagger_{l}({\bf p})b_{l}({\bf p}).
\end{equation}
The density and current operators  in the momentum space are
\begin{eqnarray}
\rho({\bf k})&=&\int d^2 xe^{-i{\bf k}\cdot{\bf x}}
\psi^\dagger({\bf x})\psi({\bf x}), \nonumber \\
{\bf j}({\bf k})&=&\int d^2 xe^{-i{\bf k}\cdot{\bf x}}
(-{i\over m})\psi^\dagger({\bf x})
({\overrightarrow{\nabla}-\overleftarrow{\nabla} \over 2}
+ie{\bf A}({\bf x})) \psi({\bf x}).
\label{cur}
\end{eqnarray}
Using a modular transformation of $\beta({\bf p})$, which is given in
Appendix A, $j^{\mu}=(\rho,{\bf j})$  becomes
\begin{equation}
j^{\mu}({\bf k})=\int_{{\rm BZ}}{d^2 p\over (2\pi)^2}
\sum_{l,l'}b^\dagger_{l}({\bf p})b_{l'}({\bf p}+a\hat{\bf k})
\langle f_{l}\vert {1\over 2}\{v^\mu ,e^{-ik\cdot\xi}\}
\vert f_{l'}\rangle
e^{{i\over4\pi}a\hat k_x(2p+a\hat k)_y}.
\label{jmu}
\end{equation}
Here, $v^\mu=(1,-\omega_c\eta,\omega_c\xi)$,
and  $\hat{k_i}=W_{ij}k_j$.
The explicit form of $\langle f_{l}\vert
e^{-ik\cdot\xi}
\vert f_{l'}\rangle$ is given in Appendix A.
The phase factor in Eq.~(\ref{jmu})  
can be written as $\exp(i\int_p^{p+a\hat{k}} A_i({\bf p}) dp_i)$,
where the path is the straight line from $p$ to $p+a\hat{k}$ and 
$A_i(p)=(p_y/2\pi,0)$. 
If we choose a non-zero $\chi({\bf p})$, then  $A_i({\bf p})$ becomes 
$A_i-\partial_i \chi$, {\it i.e.} a gauge transformation of $A_i$.  
Therefore,  $A_i({\bf p})$ is a gauge field on BZ and the 
 phase factor in Eq.~(\ref{jmu}) represents a unit flux on BZ.
We note that the simple form of $j_\mu ({\bf k})$ is due to 
our convention $\chi({\bf p})=0$.  
For later convenience, we derive the
density operator in the spatial coordinate space.
Let us define the polynomial $f_{ll'}(k)$ as
\begin{equation}
\langle f_{l}\vert e^{-ik\cdot\xi}\vert f_{l'}\rangle
=f_{ll'}(k)e^{-{a^2{\bf k}^2 \over 8\pi}}
\end{equation}
The explicit form of $f_{ll'}(k)$ is given in Appendix A.
The density operator can be written as
\begin{eqnarray}
\rho (x)
&=& {1\over a^2}\sum_{ll'}
\int_{{\rm BZ}}{d^2p\over (2\pi)^2}{d^2q\over (2\pi)^2}
b^\dagger_{l}({\bf p})b_{l'}({\bf q})
f_{ll'}({\partial \over i\partial x})
\nonumber \\
& &\times
\beta(p_x-2\pi{\tilde{y}\over a},
      p_y+2\pi{\tilde{x}\over a})
      e^{-i{\tilde{y}\over a} p_y}
\beta^* (q_x-2\pi{\tilde{y}\over a},
         q_y+2\pi{\tilde{x}\over a})
e^{i {\tilde{y}\over a} q_y} .
\end{eqnarray}
Since $\rho (x)=\psi^{\dagger}(x)\psi(x)$,
we obtain a useful identity which  relates the normalization factor 
$\beta({\bf p})$ with
the coordinate representation of the state $|l,{\bf p}\rangle$ as
\begin{equation}
a^2 \langle l,{\bf p}|x\rangle\langle x|l',{\bf q}\rangle
=f_{ll'}({\partial\over i\partial x})
\beta(p_x-2\pi{\tilde{y}\over a},
      p_y+2\pi{\tilde{x}\over a})
      e^{-i{\tilde{y}\over a}p_y}
\beta^* (q_x-2\pi{\tilde{y}\over a},
         q_y+2\pi{\tilde{x}\over a})
e^{i {\tilde{y}\over a}q_y}.
\end{equation}
As a consequence of the above identity, we obtain Eq.~(\ref{momentum}) 
for example.

The free Hamiltonian ${\cal H}_0$ is diagonal in the above basis.
However, the density operator is not diagonal with respect to
the Landau level index.
This basis, which we call the energy basis, is convenient to
describe the energy spectrum of the system.
In another basis,
${\cal H}_0$ is not diagonal and the density operator is diagonal.
This basis, which we call  the current basis, is convenient to describe
the Ward-Takahashi identity and the 
topological formula  of  Hall conductance.
There is no basis in which both the Hamiltonian and the density 
are diagonal. This is one of  peculiar features in a
magnetic field.

The current basis is constructed as follows.
Using a unitary operator, we can diagonalize
the density operator with respect to  the Landau level index.
We define the unitary operator
\begin{equation}
U^\dagger_{ll'}({\bf p})=\langle f_l\vert
e^{i p\cdot\tilde{\xi}/a-{i\over4\pi}p_xp_y}
\vert f_{l'}\rangle.
\end{equation}
By introducing a unitary transformed operator
$\tilde{b}_l({\bf p})=\sum_{l'}U_{ll'}({\bf p})b_l({\bf p})$,
the density operator is written in the diagonal form and
the current operator  becomes a simple form:
\begin{eqnarray}
\rho({\bf k})&=&\int_{{\rm BZ}}{d^2 p\over (2\pi)^2}\sum_
{l}\tilde b^\dagger_{l}({\bf p})\tilde b_{l}({\bf p}+a\hat{\bf k}).
\nonumber \\
{\bf j}({\bf k})&=&\int_{{\rm BZ}}{d^2 p\over (2\pi)^2}
\sum_{l,l'}
{\tilde b}_{l}^\dagger({\bf p})\{{\bf v}+{a\omega_c\over2\pi}(
W^{-1}{\bf p}+{a\over2}{\bf k})\}_{l,l'}
\tilde b_{l'}({\bf p}+a\hat{\bf k}).
\end{eqnarray}
$\tilde b_l$ and $\tilde b^\dagger_l$
satisfy the anti-commutation relation and  boundary condition 
\begin{eqnarray}
\{\tilde b_{l}({\bf p}),
\tilde b^\dagger_{l'}({\bf p'})\}&=&\sum_N (2\pi)^2\delta
({\bf p}-{\bf p'}-2\pi{\bf N})\Lambda_{ll'}({\bf N}), \nonumber \\
\tilde b_{l}({\bf p}+2\pi{\bf N})&=&\sum_{l'}\Lambda_{ll'}({\bf
N})\tilde b_{l'}({\bf p}),
\end{eqnarray}
where $\Lambda$ is defined by
$\Lambda({\bf N})=(-1)^{N_x+N_y}U(2\pi{\bf N})$ 
and is independent of $\bf p$. However, it shuffles the
Landau-level index.

Here we review briefly  the Ward-Takahashi identity and 
the topological formula of Hall conductance using the current basis.
We consider various interactions perturbatively.
It is convenient to define the one-particle irreducible vertex part 
$\tilde{\Gamma}^\mu$ as 
\begin{equation}
\langle j^\mu (q)\tilde{b}_l(p)\tilde{b}_{l'}^\dagger (p')\rangle 
= (2\pi)^3 \delta(p+Q-p')
 \tilde{S}_{ll_1}(p)\tilde{\Gamma}_{l_1l_2}^\mu (p,p+Q)
 \tilde{S}_{l_2l'}(p+Q),
\end{equation}
where $Q^{\mu}=(q_0,a\hat{q}_x,a\hat{q}_y)={t^\mu}_\nu q^\nu$ is a 
linear combination of $q^\mu$  and 
$\tilde{S}$ is the full propagator defined by 
\begin{equation}
\langle \tilde{b}_l(p)\tilde{b}_{l'}^\dagger (p')\rangle 
= (2\pi)^3 \delta(p-p')\tilde{S}_{ll'}(p).
\end{equation}
Between $\tilde{S}$ and $\tilde{\Gamma}^\mu$ 
the Ward-Takahashi identity is satisfied. 
The identity has crucial roles in the
following derivation of the topological formula of Hall conductance.
The Ward-Takahashi identity in this case becomes 
\begin{equation}
\tilde{\Gamma}_{\mu}(p,p)
={t^\nu}_\mu {\partial \tilde{S}^{-1}(p) \over \partial p^{\nu}}.
\label{wt}
\end{equation}
In a theory without a magnetic field, the Ward-Takahashi identity
gives a  relation that the state of the dispersion
$\epsilon(p)$ moves with the velocity
${\partial \epsilon(p) \over \partial p_i}$.
However in a magnetic field, we can not diagonalize both the current
and the energy simultaneously. Therefore, the Ward-Takahashi identity
Eq.~(\ref{wt}) does not imply the relation.

The Hall conductance is the slope of the current correlation function 
$\pi^{\mu\nu}(q)$ at the origin and is written as 
\begin{equation}
\sigma_{xy}={e^2 \over 3!}\epsilon^{\mu\nu\rho}
\partial_{\rho}\pi_{\mu\nu}(q)|_{q=0}.
\label{hallc}
\end{equation}
If the derivative $\partial_\rho$ 
acts on the vertex with the external line attached, 
its contribution becomes zero owing to the epsilon tensor.
Therefore,  the case  that the derivative acts on the bare propagator 
is  survived.
At this point it is proved that 
only diagrams of Fig.4   do contribute to 
$\sigma_{xy}$\cite{imatsu,iimt}.

\begin{figure}[h]
\centerline{
\epsfxsize=1.5in\epsffile{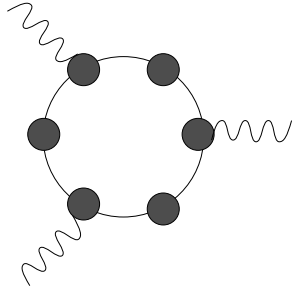}}
\caption{Feynman diagrams  which contribute to $\sigma_{xy}$.}
\end{figure}
\noindent

In Fig.4  the dark disk of the vertex  and  the dark disk of the propagator 
 are the one-particle irreducible 
vertex part  and the full propagator.
These satisfy the Ward-Takahashi identity.  
Thus, $\sigma_{xy}$ is written as a topologically invariant expression of 
the full propagator: 
\begin{equation}
\sigma_{xy}={e^2 \over h}{1\over 24\pi^2}\int_{{\rm BZ}\times S^1}
d^3 p \epsilon_{\mu\nu\rho} {\rm tr} \left(
\partial_\mu \tilde{S}^{-1}(p)\tilde{S}(p)
\partial_\nu \tilde{S}^{-1}(p)\tilde{S}(p)
\partial_\rho \tilde{S}^{-1}(p)\tilde{S}(p)\right)
\label{topological}
\end{equation}
Here, the trace is taken over the Landau level index and the $p_0$ 
integral is a contour integral on a closed path which is drawn 
in Fig.5.

\begin{figure}[h]
\centerline{
\epsfysize=1.5in\epsffile{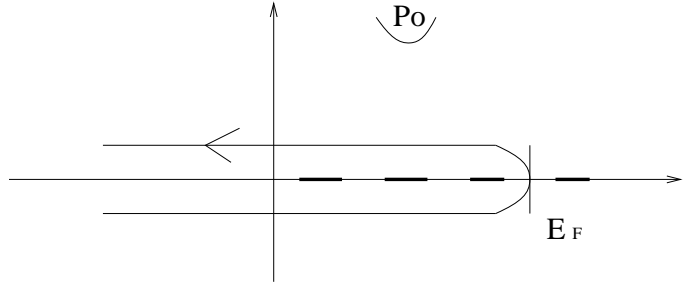}}
\caption{The contour line of $p_0$. Bold lines in the real 
$p_0$ axe represent Landau bands. The Fermi energy $E_F$ is located in 
the gap region.}
\end{figure}
\noindent

Thus, we denote $S^1$ as the integration range. 

If the Fermi energy lies in a gap region, the expression of
$\sigma_{xy}$ becomes  simple. The Coleman-Hill theorem\cite{ch} tells us that 
in this region only the lowest order diagram, {\it i.e.} the diagram of 
Fig.4 with the bare propagator and the bare vertex, does contribute to 
$\sigma_{xy}$. 
Thus, the full propagator $\tilde{S}$ can be replaced by the bare
propagator  $\tilde{S_0}$.
The integral  
${1\over 24\pi^2}\int tr(d\tilde{S_0}^{-1}\tilde{S_0})^3$ 
gives a integer value and in fact counts the number of  Landau bands 
below the Fermi energy. 
Thus, the Hall conductance is proved to
be a integer times $e^2/h$ in a gap region.

\section{Periodic Potential  and  Defect}

In this section, we apply the formalism developed in the previous
section to a system subjected to a periodic potential.
Generally, a periodic potential lifts the degeneracy of the
Landau level and the spectrum consists of Landau bands.
As we mentioned in the introduction,
Bloch electron in a strong magnetic field is of great interest
in some contexts. 
In what follows, we assume that the potential lattice  is
formed by two linear-independent basis vectors
with integer coefficients.
Such a lattice  is called a regular lattice.
A periodic potential without this property, {\it e.g.} a honeycomb
lattice potential, is not considered in this paper.
A periodic potential with this property is generally
written as
\begin{equation}
V({\bf x})=\sum_{{\bf N}\in {\bf Z}}
v({\bf x}+aN_x {\bf w}_x^{pot}+aN_y  {\bf w}_y^{pot}),
\end{equation}
where ${\bf w}_x^{pot},{\bf w}_y^{pot}$ are  basis vectors of
the potential lattice.
As is well-known, a periodic
potential problem in a magnetic field is very sensitive to the flux
penetrates a unit cell of the potential lattice.
The unit cell means a parallelogram spanned by basis vectors of the
potential lattice.
If the flux $\Phi$ is a rational multiple
of the flux quantum $\Phi_0$, that is,
\begin{equation}
t\equiv{\Phi\over \Phi_0}
={\rm Im}(\omega_y^{pot})^*\omega_x^{pot}={q\over p},
\label{ratio}
\end{equation}
each  Landau band splits into $q$ subbands.
The magnetic Brillouin zone (MBZ) is  one $q$-th of
the Brillouin zone.
In addition, it was proven that the spectrum in the
MBZ is  $p$-fold degenerate\cite{usov}. 
We re-prove this degeneracy
in a simple manner using the previously developed formalism.

When the flux is given by Eq.~(\ref{ratio}),
it is  convenient  to select  
basis vectors of the von Neumann lattice as 
$\omega_x=\omega_x^{pot}/q, \omega_y=p\omega_y^{pot}$.
The moduli of the von Neumann lattice becomes
$\tau =\tau^{pot}/pq$, where the moduli of the potential lattice
is defined  by $\tau^{pot}=-\omega_y^{pot}/\omega_x^{pot}$.
In this case, the potential energy term
in the second quantized form becomes
\begin{eqnarray}
H_{\rm pot}&=&{1\over q}\sum_{l,l'}\sum_{s=0}^{p-1}\sum_{r=0}^{q-1}
 \int d^2 x v({\bf x})
\int_{{\rm BZ}}{d^2 p \over (2\pi)^2}
b^\dagger_{l}({\bf p})b_{l'}(p_x-2\pi{r\over q},p_y) \nonumber \\
  & &\times \langle l,p_x-2\pi ({\tilde{y}\over a}+{s\over p}),
                    p_y+2\pi {\tilde{x}\over a}|{\bf x}={\bf 0}\rangle
          \langle {\bf x}={\bf 0}|l',p_x-2\pi({\tilde{y}\over a}+
                         {s\over p}+{r\over q}),
                         p_y+2\pi {\tilde{x}\over a} \rangle.
\end{eqnarray}
The eigenvalue equation reads
\begin{eqnarray}
(E-E_l)\psi_{lr}({\bf p})&=&
{1\over q} \sum_{l'}\sum_{s=0}^{p-1}\sum_{r'=0}^{q-1}  \int d^2x v({\bf x})
\langle l,p_x-2\pi ({\tilde{y}\over a}+{s\over p}+{r\over q}),
                    p_y+2\pi {\tilde{x}\over a}|{\bf 0}\rangle
                   \nonumber \\
& & \times  \langle {\bf 0}|l',p_x-2\pi({\tilde{y}\over a}+
                         {s\over p}+{r' \over q}),
                         p_y+2\pi {\tilde{x}\over a} \rangle
\psi_{l'r'} ({\bf p}),
\label{eigen}
\end{eqnarray}
where ${\bf p}$ is a momentum in MBZ, that is, 
$|p_x|\le \pi/q, |p_y|\le \pi$.
The function $\psi_{lr}({\bf p})$ is given by the following form:
\begin{equation}
\psi_{lr}({\bf p})=\psi_l (p_x-2\pi{r\over q},p_y).
\end{equation}
The equation is solved by diagonalizing 
a $Lq\times Lq$ matrix, where $L$ is the number of Landau
levels. Therefore, each Landau band splits into $q$ subbands
generally.
It is easy to see that the spectrum of the eigenvalue equation
is invariant under the  translations  $p_x \to p_x+2\pi m/p$
and $p_x \to p_x+2\pi n/q$, where $m,n$ are integers.
Since $p,q$ are relatively prime integers, the above symmetry  reads
\begin{equation}
E({\bf p})=E(p_x+2\pi{n\over pq},p_y).
\end{equation}
Thus, it is proven that  the spectrum in MBZ  is  $p$-fold degenerate.

Next, we consider  additional defects to a periodic potential.
To extract properties in such a system, let us suppose a defect 
in a periodic short-range potential.
That is, the potential is given by 
\begin{equation}
V({\bf x})=\sum_{{\bf N}\in {\bf Z}}
V_0 a^2 \delta ({\bf x}-aN_x {\bf w}_x^{pot}-aN_y  {\bf w}_y^{pot}) 
+g a^2 \delta ({\bf x}-aM_x {\bf w}_x^{pot}-aM_y  {\bf w}_y^{pot}), 
\end{equation}
for   integers  $M_x,M_y$. The second term breaks the periodicity of the 
potential.
If we neglect the second term, the eigenvalue equation becomes 
\begin{equation}
\sum_{l'r'}{V_0\over q}(D_l^\dagger ({\bf p})D_{l'}({\bf p}))_{rr'}
\psi_{l'r'}({\bf p})=(E-E_l)\psi_{lr}({\bf p}),
\label{antidot}
\end{equation}
where $p\times q$ matrix $D_l$ is given by 
\begin{equation}
(D_l({\bf p}) )_{sr}=a\langle {\bf 0}|l,p_x-2\pi ({s\over p}+{r\over q}),p_y 
\rangle  \qquad (s=0,..,p-1; r=0,..,q-1).
\end{equation}
As was discussed in our previous paper\cite{imos},
 the spectrum in the LLL consists of 
 flat bands and Hofstadter-type bands. If we take account of higher Landau 
levels and the Landau level 
mixing, flat bands still exist at  original Landau levels and also 
Hofstadter-type  bands exist. 
The wave function of the flat band at $E=E_l$ is given by 
$\psi_k=\delta_{kl} {\rm Ker}(D_l)$. 

\begin{figure}[h]
\centerline{
\epsfxsize=3in\epsffile{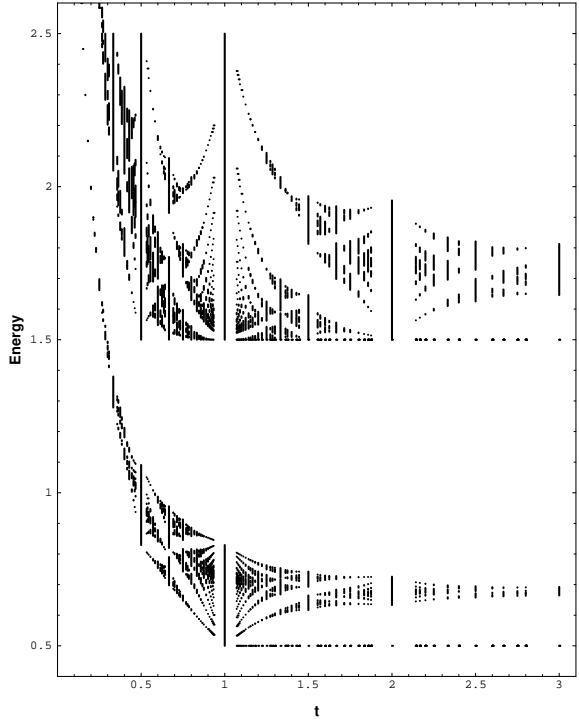}}
\caption{The spectrum of 
the periodic short-range potential problem.
The lowest two Landau bands are drawn in this figure.
The lowest five Landau bands are taken into account.
}
\end{figure}
\noindent

In Fig.6 the spectrum for the square lattice potential 
is shown. 
Here, we incorporate  lowest five 
 Landau levels and $V_0/\hbar\omega_c=0.3$ as an example. 
We observe that a self-similar pattern and large gaps above  flat bands 
exist in this figure. Hofstadter-type  bands  above flat bands  tend 
to a set of bound states as $t$ becomes infinity. 
Detailed explanation of the spectrum and the Hall conductance 
will be given in our forthcoming 
article.

Incorporating the defect leads to the following additional term in the L.H.S. 
of Eq.~(\ref{antidot}):
\begin{equation} 
g \sum_{l'm'}\int_{{\rm MBZ}} 
{d^2 q\over (2\pi)^2} (D_l^\dagger ({\bf p}))_{rs} 
( D_{l'} ({\bf q}))_{sm'} e^{iqM_x(q_x-p_x)+iS(q_y-p_y)}\psi_{l'm'} ({\bf q}),
\end{equation}
where ${\rm Mod}(M_y,p)=s$ and $(M_y-s)/p=S$. 
Apparently, flat bands are  still flat even if the defect exists. 
Furthermore, in each  subband gap of the periodic short-range potential 
problem a bound state appears rearranging  eigenstates of  
Hofstadter-type bands. 
The equation of  bound state energies is given by 
\begin{equation} 
1=g\sum_A \int_{{\rm MBZ}} {d^2 p\over (2\pi)^2} 
{|\sum_{lr} (D_l ({\bf p}))_{sr} {\psi_{lr}}^A({\bf p})e^{iqM_xp_x+iSp_y}|^2 
 \over E-E_A ({\bf p})},
\end{equation}
where  ${\psi_{lr}}^A ({\bf p})$ and $E_A ({\bf p})$ are 
the eigenfunction 
and the eigenvalue of Eq.~(\ref{antidot}). 
The  R.H.S of the above equation generally becomes infinite  
when $E$ approaches to  the upper edge of the Landau subband. 
Also, it becomes
minus infinite when $E$ approaches to  the lower edge.
Thus, there exists a solution  in each subband gap. 
The solution corresponds to a bound state trapped at the defect.  
If many defects exist in the periodic potential, 
many bound states appear in each the  subband gap. 
As the number of defects  increases, 
subband gaps tend to be  filled with bound states and 
the number of extended states decreases. 
At last states which correspond to Landau subbands 
and bound states are shuffled each other and turn into 
a set of bound states (localized states) which has a finite density of state.

\section{Quantized Hall Conductance of Bloch Electrons}

In this section, we generalize results given in section III to 
a system subjected to a periodic potential.
In what follows, we assume the periodic potential has $t=q/p$.
Thus, the each Landau band splits into $q$ subbands generally.
In this case we can not directly use the Ward-Takahashi identity 
 and thus the topological formula of the Hall conductance 
discussed in section III. 
This is because only the reduced momentum, i.e. the momentum in MBZ 
is conserved in this case. 
Therefore, we must generalize the Ward-Takahashi identity to the system 
subjected to a periodic potential first.

Let us suppose that the eigenvalue and the eigenfunction
of the one-body Hamiltonian under the periodic potential
is given by
$E_A ({\bf p}), {\psi_{lr}}^A ({\bf p})$.
Here, ${\bf p}$ is a momentum in the MBZ,
$A$ is the suffix to specify the eigenstate,
$l$ is the Landau level index
and $r=0,1,...,q-1$.
The electron propagator in the current basis is written as
\begin{eqnarray}
& &\tilde{S}_{ll'}
(p_0,p_{x}-2\pi{r \over q},p_y;p_0',p_{x}'-2\pi{r' \over q},p_y')
\nonumber \\
&=&(2\pi)^3\delta (p- p')
U_{ll_1}(p_{x}-2\pi{r \over q},p_y){\psi_{l_1r}}^A({\bf p})
{i \over p_0-E_A ({\bf p})}{{\psi_{l_2r'}}^A}^*({\bf p})
U_{l_2l'}^{\dagger}(p_{x}-2\pi{r' \over q},p_y)
\nonumber \\
&=&(2\pi)^3\delta (p-p')
\tilde{S}_{(lr)(l'r')}(p).
\label{propagator}
\end{eqnarray}
Thus the momentum in the MBZ is  conserved in the propagator.
The propagator $\tilde{S}_{(lr)(l'r')}({\bf p})$ can be regarded as a
matrix whose indices run over allowed set of ($l,r$).
Owing to the orthogonality and the completeness of $\psi$,
its inverse is given by
\begin{equation}
\tilde{S}^{-1}=U\psi {p_0-E_A \over i}\psi^{\dagger}U^{\dagger}.
\end{equation}
The one-particle irreducible vertex in the current basis is the same
as the one in the free theory, because effects of the periodic
potential are  absorbed  in the propagator.
Thus, for an infinitesimally small momentum transfer, the vertex becomes
\begin{eqnarray}
& & \tilde{\Gamma}_{ll'}^{\mu}
(p_0,p_{x}-2\pi{r \over q},p_y;p_0',p_{x}'-2\pi{r' \over q},p_y';q)
\nonumber \\
&=& -i\delta_{rr'}(2\pi)^3\delta (p+Q-p') \times \nonumber \\
& & \left( 
v^{\mu}+{1\over 2ma}[
\delta_1^{\mu} (\sum_{i=x,y}W^{-1}_{xi}(2p_i+Q_i)-W^{-1}_{xx}4\pi{r\over q})+
\delta_2^{\mu} (\sum_{i=x,y}W^{-1}_{yi}(2p_i+Q_i)-W^{-1}_{yx}4\pi{r\over q})]
\right)_{ll'}
\nonumber \\
&=&(2\pi)^3\delta (p+Q-p')
\tilde{\Gamma}_{(lr)(l'r')}^{\mu}
(p,p+Q).
\end{eqnarray}
Between $\tilde{S}$ and $\tilde{\Gamma}$, 
the  Ward-Takahashi identity is  satisfied in a generalized form. 
The Ward-Takahashi identity in this case is simply given by
\begin{equation}
\tilde{\Gamma}_{(lr)(l'r')}^{\mu}({\bf p},{\bf p})
=t^{\nu\mu}
{\partial \tilde{S}_{(lr)(l'r')}^{-1}({\bf p}) \over \partial p^{\nu}}.
\end{equation}
The  current correlation function
for infinitesimally small momentum transfers $q,q'$ is
given by
\begin{eqnarray}
\pi^{\mu\nu}(q,q')&=&
{1\over a^2}(2\pi)^3 \delta (q-q')
\int {dp_0\over 2\pi}\int_{{\rm MBZ}}{d^2 p\over (2\pi)^2}
Tr[\tilde{\Gamma}^{\mu}(p,p+Q)\tilde{S}(p+Q)
   \tilde{\Gamma}^{\nu}(p+Q,p)\tilde{S}(p)]
\nonumber \\
&=&(2\pi)^3 \delta (q-q')\pi^{\mu\nu}(q).
\end{eqnarray}
Here, the trace is taken over indices ($l,r$).
In the  expression of the Hall conductance Eq.~(\ref{hallc}), 
the term including 
$\partial_{\rho}\tilde{\Gamma}_{\mu}$ or
$\partial_{\rho}\tilde{\Gamma}_{\nu}$ are vanished
when the totally antisymmetric part is taken. The term including
$\partial_{\rho}\tilde{S}$  survives. Therefore,
using the Ward-Takahashi identity, the Hall conductance
is written as
a  topologically invariant expression of
the propagator $\tilde{S}$:
\begin{equation}
\sigma_{xy}={e^2 \over h}{1\over 24\pi^2}
\int_{{\rm MBZ}\times S^1} Tr(d\tilde{S}^{-1}\tilde{S})^3 .
\label{genetopo}
\end{equation}
Here, the ${\bf p}$ integral is taken over MBZ.
Since the integral ${1\over 24\pi^2}
\int Tr(d\tilde{S}^{-1}\tilde{S})^3 $ gives the winding number 
of the propagator which is  a integer value
under general assumptions, the Hall conductance is proved to
be a integer times $e^2/h$  whereas the filling factor is a fraction.
This formula of the Hall conductance is a generalization of
the topological formula Eq.~(\ref{topological})
 to the system subjected to a periodic
potential.

Next, we consider the relation between the winding number  and 
Chern number.  Thouless {\it et al.}  showed that 
the Hall conductance of Bloch electrons 
 in a magnetic field 
is given by the Chern number of the Bloch function.
This  formula of the Hall conductance leads to a surprising result
in a cosine potential.
That is, in the gap region of  Hofstadter bands, the Hall conductance
becomes a integer times $e^2/h$ in spite that the system
has a fractional  filling factor in the gap.
Furthermore, the Hall conductance is changed  drastically
as the Fermi energy is increased across subband gaps.
This shows each  subband carries a very large mobility.
These features of the Hall conductance in a cosine potential are also
obtained by the Streda formula\cite{streda}.
In contrast to the cosine potential 
the Hall conductance in
a periodic short-range potential is not known.
Taking account of recent antidot experiments, it is also important 
to study the Hall conductance in the periodic short-range potential. 

To solve these problems, we rewrite the generalized topological 
formula  Eq.~(\ref{genetopo}) in a more
convenient form.
In computation of Eq.~(\ref{genetopo}), there is a useful relation 
due to  Polyakov-Wiegmann\cite{pw}:
\begin{equation}
I(ST)=I(S)+I(T)+3\int Tr d(S^{-1}dSdTT^{-1}),
\label{pw}
\end{equation}
where  $I(S)$ is defined by $\int Tr(SdS^{-1})^3$. 
To utilize the relation, it is convenient to regard 
$\psi, U, S^{(0)}=i/(p_0-E_A)$ as matrices of order $Lq$. 
We regard that $U_{ll'}(p_x-2\pi r/q,p_y)$ is diagonal with respect to 
the index $r$ and that $S^{(0)}$ is  diagonal with respect to 
the index $A$. 
Therefore, the propagator $\tilde{S}$ is  regarded  as
a product of three matrices,$\psi^U=U\psi, S^{(0)}$ and $(\psi^U)^\dagger$.
Using  Eq.~(\ref{pw}), 
the generalized topological  formula leads to
\begin{equation}
\sigma_{xy}={e^2 \over h}{i\over 2\pi}\sum_{A (E_A<E_F)}
\left( \int_{{\rm MBZ}} 
 d({\psi^A}^{\dagger}d\psi^A)-2\pi i{1\over q}\right).
\label{ourchern}
\end{equation}
This formula is similar to that of Thouless {\it et al.}\cite{tknn} 
and  Kohmoto\cite{kohmoto2} 
except that there is the second term
proportional to the filling factor.
The second term comes from the matrix $U({\bf p})$. 
However, this term at last cancels with the boundary contribution 
of the first term. 
For the first term, we give a topological argument which is similar 
to the one given in \cite{kohmoto}. 
The boundary condition in BZ implies that  
the phase of the eigenfuction  varies by $2\pi$ as one follows 
 the boundary of BZ counterclockwise. 
Since BZ is divided into $q$ MBZs, this leads to  a constraint given by  
\begin{equation}
\sum_{r=0}^{q-1} \oint_{\partial {\rm MBZ}} d{\rm arg}({\psi_{lr}}^A)=2\pi.
\end{equation}
Under this constraint the configuration which is singular in 
${\psi^A}^\dagger d\psi^A $ at some ${\bf p}$ may appears for a 
band $A$. We call this the  vortex. 
If the vortex appears, ${\psi}^\dagger d\psi$ 
is ill-defined in the entire MBZ.   
Let us  assume there is a vortex at ${\bf p}={\bf p}_c$.
We remark that 
there is  a gauge degree of freedom in  the eigenfunction. 
That is, the overall phase factor of the eigenfunction $\psi^A ({\bf p})$  
is not determined.
As is well known in the description of monopole, 
 we can define ${\psi}^\dagger d\psi$ 
in the entire MBZ using the gauge degree of freedom.
In order to define it, we divide  MBZ
into two regions. One region (${\rm MBZ}_1$) does not include ${\bf p}_c$ 
 and has a boundary   of  MBZ.
The other region (${\rm MBZ}_2$) includes ${\bf p}_c$. 
The boundary between  two regions is a closed line around 
 ${\bf p}_c$.
In ${\rm MBZ}_2$ using the gauge degree of freedom,  
 we choose  a different phase assignment which has no vortex.
It is not necessary to satisfy the boundary condition 
when we extrapolate the wave function  to ${\rm MBZ}_1$. 
Thus,  $\psi^\dagger d\psi$  is completely defined  in the entire MBZ.
 However,
there is a phase mismatch at the boundary
between  two regions. 
 That is, the eigenfunction  $\psi^{(1)}$ in ${\rm MBZ}_1$ 
and the eigenfunction $\psi^{(2)}$ in ${\rm MBZ}_2$  are  related as
$\psi^{(2)}=e^{i\theta}\psi^{(1)}$ at the boundary.
For the integral $\int_{{\rm MBZ}_i}d(\psi^\dagger d\psi)  (i=1,2)$,
we can use the Stokes theorem and obtain 
\begin{equation}
\int_{{\rm MBZ}} d(\psi^\dagger d\psi)=
\int_{\partial {\rm MBZ}} \psi^\dagger d\psi+ i\int_{C} d\theta,
\end{equation}
where $C$ is the closed line around ${\bf p}_c$.
The first term in R.H.S. of the above equation is equal to $2\pi i/q$
due to  the normalization and the boundary condition 
 of the eigenfunction $\psi$.
The second term gives 
$2\pi i$ times a integer. This integer is called the Chern number.
Therefore, starting  from the generalized topological formula
we obtain the Chern number for the Hall conductance:
\begin{equation}
\sigma_{xy}=-{e^2 \over h}{1\over 2\pi}
\sum_{A (E_A<E_F)}
\oint_C d\theta^A. 
\end{equation}

To compare our result with the one obtained by  Thouless {\it et al.}
\cite{tknn}, we consider the correspondence between the Bloch function 
$u_{{\bf k}}$ and the eigenfunction $\psi_{lr}$ in the von Neumann lattice 
basis. 
In terms of the Bloch function the  energy eigenstate can be written as 
\begin{equation}
u_{{\bf k}}({\bf x})e^{i{\bf k}\cdot{\bf x}}.
\label{bloch}
\end{equation}
On the other hand, in the von Neumann lattice basis 
the  energy eigenstate can be written as
\begin{equation}
{1\over \sqrt{q}}\sum_{l,r}\psi_{lr}({\bf p})
\langle {\bf x}|l,p_x-2\pi{r\over q},p_y \rangle.
\label{our}
\end{equation}
Both functions satisfy the same boundary condition (except for a
constant phase) and normalization in the magnetic
unit cell (MUC) if the momentum ${\bf k}$ is replaced by the momentum   
${\bf p}$ with the rule 
$a{\bf k}\cdot {\bf w}_x^{pot}=qp_x,\quad  
pa{\bf k}\cdot {\bf w}_y^{pot}= p_y$
in a $t=q/p$ problem. 
Thus Eq.~(\ref{bloch}) can be identified with Eq.~(\ref{our}).
The  Chern number obtained by  Thouless {\it et al.} is given by 
\begin{equation}
n_C={1\over 2\pi i}\int_{{\rm MBZ}} dA,
\label{cherntknn}
\end{equation}
where $A$ is the connection 1-form written as 
\begin{equation}
A({\bf k})=\int_{{\rm MUC}}d^2 x  u_{{\bf k}}^* ({\bf x})d_{{\bf k}}
 u_{{\bf k}}({\bf x}). 
\end{equation}
Substituting Eq.~(\ref{our}) into Eq.~(\ref{cherntknn}), we obtain 
\begin{equation}
n_C={1\over 2\pi i}\int_{{\rm MBZ}}d(\psi^\dagger d\psi) -{1\over q}.
\end{equation}
This Chern number is nothing but the one in Eq.~(\ref{ourchern}).
Therefore, the winding number of the propagator  is reduced to 
the Chern number of Thouless {\it et. al.} if  interactions are neglected.

So far, we neglected interactions and consider only a periodic potential.
However, in an interacting system subjected to the periodic potential 
 the winding number formula of the 
Hall conductance can be applied in the following way. 
In this system  
we can treat the periodic potential nonperturbatively as discussed above 
and treat  interactions perturbatively. 
Thus, the propagator Eq.~(\ref{propagator}) replaces the ``bare'' propagator. 
Since the Ward-Takahashi identity and related theorems still hold, 
results obtained in Sec.3 can be also applied in this system.
Therefore, the Hall conductance in a gap region is written by 
the winding number of the propagator Eq.~(\ref{propagator}). Furthermore,  
the winding number is the same as the Chern number obtained by Thouless 
{\it et al.}  as discussed above.

Finally we calculate the Hall conductance in the system of a periodic 
short-range potential.
To calculate the Chern number for a general periodic potential is a subtle
problem. However, we can calculate the Chern number for some analytically
solved potentials. Let us first consider the  
potential with $t=1/p$.
This is the  case that each Landau band does not split into subbands.
In this case, the ${\bf p}$-dependence of $\psi_l({\bf p})$ can be
set as $\beta ({\bf p})/|\beta ({\bf p})|$ which satisfies the boundary
condition. Apparently, this has a vortex at ${\bf p}={\bf 0}$. Thus, we must
choose another gauge for the wave function around ${\bf p}={\bf 0}$.
A simple choice is 1, that is, there is no ${\bf p}$-dependence  around
${\bf p}={\bf 0}$. Since $\oint_C d{\rm arg}\beta ({\bf p})=2\pi $,
the Chern number of  each Landau band is equal to 1.
Thus, the Hall conductance is $e^2/h$ times  the number
of Landau bands  below the Fermi energy as expected.

Next we consider a $t=2$ case. A periodic potential
with $t=2$ can be solved analytically if the number of
truncated  Landau levels is one or two. Let us consider
the periodic short-range potential in  the lowest Landau level  approximation.
In this case the spectrum consists of a flat band and a lifted band. 
The wave function of the  flat band satisfies the equation
\begin{equation}
\beta^* ({\bf p})\psi_0({\bf p})+\beta^* (p_x-\pi,p_y)\psi_1({\bf p})=0.
\end{equation}
Here, $\psi_k ({\bf p})=\psi(p_x-k\pi,p_y)$ and 
MBZ is the region $|p_x|\le \pi/2 , |p_y|\le \pi$. 
The following choice of the phase is consistent with the boundary condition
of $y$-direction:
\begin{equation}
  \left(
  \begin{array}{c}
  \psi_0({\bf p}) \\
  \psi_1({\bf p})
  \end{array}
  \right) ={1\over \sqrt{\alpha ({\bf p})+\alpha (p_x-\pi,p_y)}}
   \left(
   \begin{array}{c}
   |\beta (p_x-\pi,p_y)|e^{i{\rm arg}(\beta ({\bf p}))+ip_x} \\
   -|\beta ({\bf p})|e^{i{\rm arg}(\beta (p_x-\pi,p_y))+ip_x}
   \end{array}
   \right).
\end{equation}
At ${\bf p}={\bf 0}$ the upper component of the above eigenfunction 
has a vortex. 
Therefore, we must choose another patch around  ${\bf p}={\bf 0}$.
A possible choice of the eigenfunction around   ${\bf p}={\bf 0}$ 
is given by 
\begin{equation}
  \left(
  \begin{array}{c}
  \psi_0({\bf p}) \\
  \psi_1({\bf p})
  \end{array}
  \right) ={1\over \sqrt{\alpha ({\bf p})+\alpha (p_x-\pi,p_y)}}
   \left(
   \begin{array}{c}
    \beta^* (p_x-\pi,p_y)  \\
    -\beta^* ({\bf p})
   \end{array}
   \right).
\end{equation}
Thus there is a phase mismatch 
$\theta=-{\rm arg}\beta ({\bf p})-{\rm arg}\beta (p_x-\pi,p_y)$ 
at the boundary. 
Its integral along the boundary is $-2\pi$. Thus, we conclude that 
the Hall conductance of the flat band is $e^2/h$. 
The eigenvalue equation for the lifted band becomes 
\begin{equation}
\beta (p_x-\pi,p_y)\psi_0({\bf p})-\beta({\bf p})\psi_1({\bf p})=0. 
\end{equation}
Its solution which is consistent with the boundary condition is 
\begin{equation}
  \left(
  \begin{array}{c}
  \psi_0({\bf p}) \\
  \psi_1({\bf p})
  \end{array}
  \right) ={1\over \sqrt{\alpha ({\bf p})+\alpha (p_x-\pi,p_y)}}
   \left(
   \begin{array}{c}
    \beta ({\bf p})  \\
    \beta (p_x-\pi,p_y)
   \end{array}
   \right).
\end{equation}
This solution has no vortex.
Thus, the Hall conductance of the lifted band is $0$. 
We can see that the Wannier state constructed from the flat  band 
is relatively spread  than the one from the lifted band.

In the case that  
  $t$ is an integer larger than $2$, 
the spectrum also consists of a flat band and a lifted band in the 
lowest Landau level. 
The eigenfunction of the lifted band 
\begin{equation}
\psi_k({\bf p})={\beta (p_x-2\pi {k\over t},p_y)\over 
\sqrt{\sum_{k=0}^{t-1}\alpha  (p_x-2\pi {k\over t},p_y)}}
\end{equation}
satisfies the boundary condition and has no vortex in the MBZ. 
Thus, the Hall conductance of the lifted band is $0$. 
This  result is reasonable because 
the lifted band becomes a set of bound states  and thus has no mobility 
as $t$ becomes infinity. Since it is proved  that 
the Hall conductance of the Landau band is $e^2/h$, the Hall conductance 
of the flat band is  $e^2/h$. 
These results are irrespective to the lattice structure of the periodic 
short-range potential.

\section{SUMMARY AND DISCUSSION}

We constructed three sets of one-particle states 
based on  the von Neumann lattice 
in a gauge-independent manner. The first one  is the set  of  
coherent sates which are localized on the sites of the von Neumann 
lattice and not orthogonal. 
The second one  is the set  of  momentum  states which are  extended, 
orthogonal and normalized with $\delta$-function.  
The third one is the set of  Wannier  states  which is  localized 
on  sites of the von Neumann lattice and is  orthonormal.
Using the second one  we developed a field-theoretical formalism 
of a quantum Hall system and showed  the Ward-Takahashi identity and 
the topological formula of the Hall conductance. 
This formalism  was applied to a system subjected to a  periodic potential.
The $p$-fold degeneracy of the spectrum in a $t=q/p$ problem is easily 
proved using this formalism. 
The generalized topological formula for the Hall conductance of Bloch 
electron was obtained. 
The formula is written as a winding number of the propagator.
Relation between the winding number and the Chern number was clarified.
As an example we calculated the quantized Hall conductance 
in a periodic short-range potential. 

In this paper we study the system subjected to a periodic potential 
 using the set of momentum states. 
We believe that our formalism is also useful to study
Coulomb interaction in  the quantum liquid phase\cite{im}
and the Wigner crystal phase\cite{crystal} of  the quantum Hall system. 
In the Wigner crystal phase, a triangular lattice 
potential is generated from the mean field of the charge density. 
In this case, the filling factor $\nu$ is related to $t$ with 
$\nu=1/t$. 
If the mean field contributes to the short-range potential with   
positive  coefficient, the Fermi energy lies 
in the large energy gap at $\nu=1-1/t$. 
That is, the flat band of the lowest Landau level in Fig.6 is 
completely filled.
Then, $\nu=1/2$ state can be a self-consistent solution which has a 
large energy gap in contrast to the quantum liquid phase 
in which gap closes at $\nu=1/2$. 
If the mean field contributes to the short-range potential with   
negative  coefficient, the Fermi energy lies 
in the large energy gap at $\nu=1/t$. 
That is, the lifted band is completely filled.
Then, all states of $\nu<1$ can be a self-consistent solution which has a 
large energy gap. 
Since the Hall conductance of the lifted band is zero, the ground state 
in this approximation is close to the crystal.

Although the short-range potential is not a good approximation for 
the Coulomb interaction, the mean field treatment starting from the 
short-range potential may give some insights to the Wigner crystal 
phase of  the quantum Hall system.

\acknowledgements
This work was partially supported by the special Grant-in-Aid
for Promotion of Education and Science in Hokkaido University
provided by the Ministry of Education, Science, Sports and Culture,
the Grant-in-Aid for Scientific Research on Priority Area (Physics of CP 
violation) (Grant No. 10140201), and the
Grant-in-Aid for International Scientific Research (Joint Research
 Grant No. 10044043) from
the Ministry of Education, Science, Sports and Culture, Japan.

\appendix
\section{}
The properties of functions used in this paper are given here.
A theta function $\vartheta_1(z\vert\tau)$ is defined by
\begin{equation}
\vartheta_1(z\vert\tau)=2\sum_{n=0}^{\infty}(-1)^n e^{\pi\tau i(n+
{1\over2})^2}\sin[\pi z(2n+1)].
\end{equation}
Under the modular transformation\cite{gsw}, 
$\beta({\bf p}\vert\tau)$ defined in Eq.~(\ref{beta}) transforms
as follows:
\begin{eqnarray}
\beta(p_x,p_y\vert {a\tau+b\over c\tau+d})=&\eta&\sqrt{c\tau+d\over
\vert c\tau+d\vert}e^{{i\over4\pi}(abp_y^2+
2bcp_xp_y+cdp_x^2)}\nonumber\\
&\times&\beta(dp_x+bp_y,cp_x+ap_y\vert\tau),
\end{eqnarray}
$$
\left(\begin{array}{cc}
a&b\\
c&d\\
\end{array}
\right)\in SL(2,{\bf Z}),
$$
where $\eta$ is an eighth-root of unity.
$\alpha({\bf p})$ defined in Eq.~(\ref{alpha}) obeys the periodic
boundary condition.
$G=\ln\alpha$ is a Green function of the Laplacian operator
on a torus\cite{fms}, that is,
\begin{equation}
\triangle G(z)=2\pi\delta^{(2)}(z)-{4\pi\over{\rm Im\tau}},
\label{green}
\end{equation}
where
$$
\delta^{(2)}(z)=\sum_N\delta^{(2)}(z+N_x+\tau N_y),
$$
$z=(p_x+\tau p_y)/2\pi$, and $\triangle=4\partial_z\bar\partial_z$.
As mentioned in Sec.~II, the isolated singularity of the delta-function
at ${\bf p}=0$ in Eq.~(\ref{green}) is not involved in our theory.
The factor $\langle f_l\vert e^{-ik\cdot\xi}
\vert f_{l'}\rangle$  which appears in $\rho({\bf k})$ is written by
\begin{eqnarray}
\langle f_l\vert e^{-ik\cdot
\xi}\vert f_{l+n}\rangle=&\sqrt{l!\over (l+n)!}\left(
-{a(k_x+ik_y)\over\sqrt{4\pi}}\right)^n& e^{-{(ak)^2\over8\pi}}
\nonumber\\
&\times L^{(n)}_l({(ak)^2\over4\pi}),&
\end{eqnarray}
where $L^{(n)}_l$ is the associated Laguerre polynomial.


\end{document}